\pdfoutput=1
\documentclass[a4paper,USenglish,cleveref,autoref, thm-restate]{lipics-v2021}

\newif\ifarxiv
\arxivtrue

\ifarxiv
\hideLIPIcs
\else
\relatedversion{TODO}
\fi

\usepackage{amssymb}
\usepackage{amsmath}
\usepackage{graphicx}
\usepackage{wrapfig}
\usepackage{xspace}
\usepackage{url}
\usepackage{color}
\usepackage{comment}
\usepackage{thmtools, thm-restate}

\ifarxiv
\usepackage[appendix=inline]{apxproof}
\else
\usepackage[appendix=strip]{apxproof}
\fi

\newtheorem{fact}[theorem]{Fact}

\newcommand{\namedref}[2]{\hyperref[#2]{#1~\ref*{#2}}}
\newcommand{\sectionref}[1]{\namedref{Section}{#1}}

\newcommand{\theoremref}[1]{\namedref{Theorem}{#1}}
\newcommand{\defref}[1]{\namedref{Definition}{#1}}
\newcommand{\figref}[1]{\namedref{Figure}{#1}}
\newcommand{\lemmaref}[1]{\namedref{Lemma}{#1}}
\newcommand{\propositionref}[1]{\namedref{Proposition}{#1}}

\newcommand{\corollaryref}[1]{\namedref{Corollary}{#1}}

\newcommand{\equalityref}[1]{\hyperref[#1]{Eq.~\ref*{#1}}}
\newcommand{\inequalityref}[1]{\hyperref[#1]{Inequality~(\ref*{#1})}}

\newcommand{\stepref}[1]{\hyperref[#1]{Step~\ref*{#1}}}

\def\boldhead#1:{\par\vskip 7pt\noindent{\bf #1:}\hskip 10pt}
\def\ithead#1:{\par\vskip 7pt\noindent{\it #1:}\hskip 10pt}
\def\inline#1:{\par\vskip 7pt\noindent{\bf #1:}\hskip 10pt}

\long\def\comment #1\commentend{}
\long\def\commfull #1\commend{#1}
\long\def\commabs #1\commenda{}
\long\def\commtim #1\commendt{#1}
\long\def\commb #1\commbend{}
\def\blackslug{\hbox{\hskip 1pt \vrule width 4pt height 8pt
    depth 1.5pt \hskip 1pt}}
\def\QED{\quad\blackslug\lower 8.5pt\null\par}

\def\inQED{~~~~~\quad\blackslug\lower 8.5pt\null}
\def\Proof{\par\noindent{\bf Proof:~}}
\def\ProofOf#1#2{\par\noindent{\bf Proof of #1 \ref{#2}:~}}

\long\def\PPP#1{\noindent{\bf Proof:}{ #1}{\quad\blackslug\lower 8.5pt\null}}

\long\def\denspar #1\densend
{#1}
\renewcommand{\paragraph}[1]{\par\noindent\textbf{#1}}

\newcommand{\Set}[1]{\left\{ #1 \right\}}

\newcommand{\Range}[1]{\left\{1,\ldots, #1 \right\}}
\newcommand{\ceil}[1]{\left\lceil #1 \right\rceil}

\newcommand{\ignore}[1]{}

\def\DEF{\stackrel{\rm def}{=}}

\def\cA{{\cal A}}

\def\cC{{\cal C}}

\def\cP{{\cal P}}

\newif\ifnotesw\noteswtrue%
   {\ifnotesw\marginpar[\hfill\(\top\)]{\(\top\)}\fi}%
   {\ifnotesw\marginpar[\hfill\(\bot\)]{\(\bot\)}\fi}

\newcommand{\mnote}[1]%
    {\ifnotesw\marginpar%
        [{\scriptsize\begin{minipage}[t]{\marginparwidth}
        \raggedleft#1%
                        \end{minipage}}]%
        {\scriptsize\begin{minipage}[t]{\marginparwidth}
        \raggedright#1%
                        \end{minipage}}%
    \fi}

\def\Plus{\hbox{\raise 0.3ex\hbox{\tiny +}}} %

\def\mtoday{\ifcase\month\or
  January\or February\or March\or April\or May\or June\or July\or
  August\or September\or October\or November\or
  December\fi\space\number\year}

\newcommand{\Part}{\mathrm{level}}

\allowdisplaybreaks

\nolinenumbers

\def\Alg{$\mathcal{A}$\xspace}%
\def\alg{\mathcal{A}}
\def\cB{{\cal B}}
\def\C{\overline{c}}
\def\Cost{\textsf{cost}}

\def\Opt{\Off}

\def\cC{{\cal C}}
\def\cP{{\cal P}}
\def\Cost{\mathsf{cost}} 
\def\Off{\mathrm{OPT}}

\def\recx{\texttt{recx}}
\def\recolor{\texttt{recolor}}
\def\promote{\texttt{promote}}

\title{Colorful Vertex Recoloring of Bipartite Graphs}

\author{Boaz {Patt-Shamir}}{School of Electrical Engineering, Tel Aviv University}{boaz@tau.ac.il}
{https://orcid.org/0000-0001-8398-8218}
{This research was supported by the Israel Science Foundation, grant No.~1948/21.}

\author{Adi Ros\'en}{CNRS and Universit\'e Paris Cit\'e}
{adiro@irif.fr}
{}
{Research partially supported by CNRS grant IEA ALFRED and ANR grant AlgoriDAM.}

\author{Seeun William Umboh}{School of Computing and Information Systems, The University of Melbourne, Australia \and ARC Training Centre in Optimisation Technologies, Integrated Methodologies, and Applications (OPTIMA)}
{william.umboh@unimelb.edu.au}
{https://orcid.org/0000-0001-6984-4007}
{Funded by CNRS grant IEA ALFRED and by the Australian Government
  through the Australian Research Council DP240101353.}

\authorrunning{B.~Patt-Shamir, A.~Ros\'en and S.\,W.~Umboh}

\Copyright{Boaz Patt-Shamir, Adi Ros\'en and Seeun William Umboh}


\relatedversion{}
\category{}


\EventEditors{Olaf Beyersdorff, Micha\l{} Pilipczuk, Elaine Pimentel, and Nguyen Kim Thang}
\EventNoEds{4}
\EventLongTitle{42nd International Symposium on Theoretical Aspects of Computer Science (STACS 2025)}
\EventShortTitle{STACS 2025}
\EventAcronym{STACS}
\EventYear{2025}
\EventDate{March 4--7, 2025}
\EventLocation{Jena, Germany}
\EventLogo{}
\SeriesVolume{327}
\ArticleNo{49}

\begin{document}

\maketitle

\begin{abstract}
We consider the problem of vertex recoloring: we
are given $n$ vertices with their initial coloring, and edges arrive
in an online fashion. The algorithm is required to maintain a valid
coloring by means of vertex recoloring, where recoloring a vertex
incurs a cost. The problem abstracts a scenario of job placement in
machines (possibly in the cloud), where vertices represent jobs,
colors represent machines, and edges represent ``anti affinity''
(disengagement) constraints. online coloring in this setting is a hard problem, and only
a few cases were analyzed.  One family of instances which is fairly
well-understood is bipartite graphs, i.e., instances in which two
colors are sufficient to satisfy all constraints.  In this case it
is known that the competitive ratio of vertex recoloring is
$\Theta(\log n)$.

In this paper we propose a generalization of the problem, which
allows using additional colors (possibly at a higher cost), to
improve overall performance.  Concretely, we analyze the simple case
of bipartite graphs of bounded largest \emph{bond} (a bond of a connected
graph is
an edge-cut that partitions the graph into two
connected components).  From the upper bound perspective, we propose
two algorithms. One algorithm exhibits a trade-off for the
uniform-cost case: given $\Omega(\log\beta)\le c\le O(\log n)$ colors, the algorithm
guarantees that its cost is at most $O(\frac{\log n}{c})$ times the
optimal offline cost for two colors, where $n$ is the number of vertices
and $\beta$ is the size of the largest bond of the graph. The other algorithm is designed
for the case where the additional colors come at a higher cost,
$D>1$: given $\Delta$ additional colors, where $\Delta$ is the
maximum degree in the graph, the algorithm guarantees a competitive
ratio of $O(\log D)$.  From the lower bounds viewpoint, we show that if
the cost of the extra colors is $D>1$, no algorithm (even
randomized) can achieve a competitive ratio of $o(\log D)$. We also
show that in the case of general bipartite graphs (i.e., of
unbounded bond size), any deterministic online algorithm has competitive
ratio $\Omega(\min(D,\log n))$.
\end{abstract}

\clearpage
\setcounter{page}{1}

\section{Introduction}
\paragraph{Motivation.}
In the cloud, jobs are placed on machines according to multiple
criteria. Sometimes, during the execution of a job, it turns out
that the job is in conflict with another job, in the sense that the
two jobs must not run on the same machine. Such conflicts may arise
due to limited resources on a single machine, or due to security
considerations, or other reasons. Whenever such a conflict between two
jobs located in the same machine is detected, the system needs to
migrate one of the conflicting jobs (or both) so as to separate
them. This situation was abstracted in \cite{AMPT-22} as the ``vertex
recoloring'' (or ``disengagement'') problem. Generally speaking, the problem
is stated as an online problem, where initially we are given $n$
vertices (representing jobs) colored in $k$ colors (representing
machines). Edges (representing conflicts) arrive online, and the
algorithm is asked to output a valid vertex coloring after the arrival
of each edge. Recoloring a vertex incurs a cost, and the cost incurred by an
algorithm is the total cost of vertex 
recoloring events by the algorithm. As
usual with online problems, the performance measure is the competitive ratio, namely the
worst-case ratio between the cost paid by the online algorithm in
question and the best cost possible (offline).

In this paper we seek to generalize the problem so as to make it both
richer theoretically and more realistic practically. Our starting
point is the following. Recalling the motivating scenario
of job migration as outlined above, we note that in many cases, it is
possible to acquire 
more machines on which jobs may be executed. Therefore, we are
interested in understanding what are the consequences of 
allowing  the online algorithm to use more colors. Our reference cost
is still  the optimal cost when using the initial $k$ colors, 
but now we allow the
online algorithm to use more colors.

The idea is not new: in fact, it was used in the very first paper on competitive analysis~\cite{SleatorT-85} (for cache
size). Later work called this type of
comparison ``resource augmentation'' (see, e.g.,~\cite{Borodin-ElYaniv}). However, in this paper, still motivated by
the job migration scenario, we propose an additional generalization: 
we also
consider the case in which the additional colors come at a greater
cost. We are not aware of any previous study of such a model.%
\footnote{In paging, this could be interpreted as follows: The offline
  solution uses a cache of size $k$, while the online solution has the
  cache of size $k$ and an additional cache whose \emph{replacement
  cost may
    be greater.}
}
We believe that this approach, which we call \emph{weighted} resource
augmentation, may be appropriate in other scenarios as well.

\paragraph{Our results.}
 The problem of vertex recoloring is natural and applicable in many
situations, but one has to bear in mind that it entails the problem of
graph coloring, which is notoriously hard, computationally
speaking. %
Following the approach of \cite{AMPT-22}, we focus
on polynomially-solvable instances of coloring. 
Specifically, in this
paper we consider bipartite graphs, i.e., we assume that after all edges
have arrived, the graph is 2-colorable. In \cite{AMPT-22} it was shown
that the competitive ratio of online vertex recoloring is $\Theta(\log
n)$ in the case of bipartite graphs, where $n$ is the number of vertices. We study the effect of using
more colors in this case. In the following, the competitive ratio is w.r.t.\ the 2-color optimal solution. For general bipartite graphs and deterministic algorithms it turns out 
that extra colors do not really help:

\begin{restatable}{theorem}{thmBond}
  \label{thm:lb-bond}
  Let $\cal I$ be the set of instances of recoloring with two basic
  colors and $n$ special colors, where recoloring by a basic color costs 1 and recoloring by a
  special color costs $D$. Then for every deterministic online recoloring
  algorithm
  there is an instance in $\cal I$ with competitive ratio
  $\Omega(\min\{\log n, D\})$.
\end{restatable}

(We note that competitive ratio $D$ is trivial to achieve when given $n$
special colors. See \lemmaref{lem:cr=D}.) 
We therefore restrict our attention to a subclass of bipartite graphs,
namely bipartite graphs of bounded \emph{bond} size. 
A bond of  a connected graph is a set of edges whose removal partitions the
graphs into two connected components (alternatively, a minimal edge cut, cf.~\cite{10.5555/1481153}, and see \defref{def:bond}).
Our main result in this paper is the following:
\begin{theorem}
  \label{thm:cost-D}
Suppose that the final graph is bipartite, with all vertex degrees at most
$\Delta$,  and with bond size at most $\beta$ for each  of its
connected components.  Then there is a deterministic algorithm
  that uses $\Delta$ special colors of cost at most $D$, whose
  competitive ratio is $O(\log D+\beta^2)$.
\end{theorem}

We note that in acyclic graphs, the largest bond size is $1$. The following result shows that
the competitive ratio of \theoremref{thm:cost-D} cannot be improved 
even if we restrict
the instances to have largest bond size 1.
\begin{restatable}{theorem}{thmLBpaths}
  \label{thm:lb-paths}
  Consider instances in which the final graph is a collection of
  paths, and there is an infinite set of special colors, each of cost $D>1$.
  The competitive ratio of any (possibly randomized) recoloring
  algorithm in this class of instances is $\Omega(\log D)$.
\end{restatable}

On the other hand, if $D=1$ (i.e., 
the cost of all colors is $1$), another algorithm provides a way to reduce
the competitive ratio at the price of using more colors.
\begin{restatable}{theorem}{tradeoffthm}
\label{thm:tradeoff}
Suppose that the final graph is bipartite with bond at most $\beta$. Then for any given
$\frac1{2\log\beta+3}<\epsilon\le1$ there is a deterministic online algorithm that uses
$O(\epsilon\log n)$ special colors of unit cost, and guarantees
competitive ratio $O(\epsilon^{-1})$.
\end{restatable}

\paragraph{Our techniques.}
Our algorithms use Algorithm~1 of \cite{AMPT-22}, denoted $\alg$ in
this paper, as a black box.
To facilitate our algorithms, however,
we need to develop a refined analysis
of Algorithm~$\alg$. In \lemmaref{lem:moderate}
we prove a tighter upper bound
on the cost of $\alg$ when applied to an input sequence 
that satisfies a certain condition
(``moderate sequences,'' cf.~\defref{def:moderate}).
Using this bound, the main idea in our algorithms is to 
filter the input sequence so as to make it moderate,
and to feed that sequence to  $\cA$  for simulation (which
we know to have
good performance). The difference between our algorithms is what to do
with the other edges, and how to determine the color of
vertices which may be affected by the filtering.

The approach taken by our Algorithm $\cB$
(whose performance is stated \theoremref{thm:cost-D}), 
is to use special colors, 
essentially in greedy fashion,
to resolve conflicts that involve edges that were 
rejected from simulation due to the moderate sequence condition. 
We show that this approach can guarantee $O(\log D)$ competitiveness
while using $\Delta$ additional colors.
The other approach, taken by our Algorithm $\cC$, is
to have a hierarchical set of instantations (simulations) of $\alg$.
If an edge is rejected by one instantation, 
one of its endpoints is sent to the next instantiation
in the hierarchy, which uses
a new set of colors; and if this fails,
we send that vertex to the next instantiation etc.
We show that the number of edges in this hierarchy of simulations
decreases exponentially as a function of a parameter
of the ``moderation'' of the
input sequence, which allows us to prove \theoremref{thm:tradeoff}.

We note that the largest bond size proves to be a very 
useful graph parameter, as it measures, in some precise sense,
how far a graph is from a tree. Our \lemmaref{lem:bond}
gives us a tool which may be handy in other contexts as well.

\paragraph{Related work.}
As mentioned above, the problem of vertex recoloring was introduced 
in \cite{AMPT-22}, where vertices have weights, and the cost of recoloring a vertex is its weight. It is shown that for bipartite
graphs, competitive ratio of $O(\log n)$ can be achieved by a deterministic online algorithm, and that no randomized algorithm has
competitive ratio $o(\log n)$. Tight bounds are also presented for $(\Delta+1)$ coloring for randomized and deterministic algorithms where $\Delta$ denotes the maximal degree in the graph. Recently, Rajaraman and Wasim~\cite{RajaramanW24} considered the capacitated setting where there is a bound $B$ on the number or weight of vertices in each color. They give ``traditional'' resource-augmented algorithms%
: the algorithms are allowed to violate the capacity bound by a $(1+\epsilon)$ where $\epsilon$ is an arbitrarily small constant. They also study the $(1+\epsilon)$-overprovisioned setting where the the algorithm is allowed $\Delta$ colors and the maximum degree of the graph is bounded by $(1-\epsilon)\Delta$.

Recoloring (or coloring with recourse) has been considered previously
in the context of dynamic data structures %
\cite{bosek_et_al,barba_et_al,solomon_and_wein,KashyopNNP23}.
In these papers, no initial coloring is given and the competitive ratio is not analyzed; their measure of performance is the absolute number of recolorings. 
Competitive analysis is implicit in
\cite{barba_et_al,solomon_and_wein} which is bicriteria: the arrival
model is similar to ours but the final graph may be arbitrary, and the
goal is to minimize both the update time and the number of colors
used.  In this line of work, it is assumed that the algorithm has
access to an oracle that can be queried about the chromatic number of
a graph.  The best general result is due to Solomon and Wein~\cite{solomon_and_wein},
who give a deterministic algorithm with $O(d)$ amortized running time
using $O\left(\frac{\log^3 n}{d}\chi(G)\right)$ colors. Henzinger et al.~\cite{hnw20} give better results for
bounded arboricity graphs.

Although the largest bond and maximum cut of a graph may seem superficially similar, they are quite different, both in value and complexity.
For example, finding the largest graph bond in bipartite 
graphs is NP-hard~\cite{bond}, but max-cut is trivial in bipartite graphs, and it is polynomially computable in planar graphs  \cite{Hadlock75}; any tree has a max-cut of size $\Omega(n)$ but largest bond size $1$.

\paragraph{Paper organization.}
The remainder of this paper is organized as follows.
In \sectionref{sec:prel}, we formalize the problem and introduce some notation.
In \sectionref{sec:non-uniform}, we prove our main result
\theoremref{thm:cost-D}. 
In \sectionref{sec:uniform}, we consider the uniform case and prove
\theoremref{thm:tradeoff}. 
\ifarxiv
In \sectionref{sec:lb},
we prove our lower bounds \theoremref{thm:lb-paths} and \theoremref{thm:lb-bond}. In \sectionref{sec:A-non-uniform}, we present a slightly improved version of Algorithm $\cB$.
\else
Additional material is included in the full version: 
we prove our lower bounds \theoremref{thm:lb-paths} and \theoremref{thm:lb-bond}, and present a slightly improved version of Algorithm $\cB$.
\fi
A short conclusion is presented in \sectionref{sec:conc}.

\section{Problem Statement and Notation}
\label{sec:prel}

\paragraph{Problem statement.}
We consider the following model. Initially we are given a palette $P$
of $k$ \emph{basic} colors, and a superset $P^*\supseteq P$ of
colors. The extra colors in $P^*\setminus P$ are called \emph{special}
colors.
 Each color $j$ is assigned a cost
$\Cost:P^*\to\mathbb{R}^+$, such that $\Cost(j)=1$ for all basic colors,
and $1\le\Cost(j)\le D$ for all special colors, where $D\ge1$
is some given parameter.
We are also given a set of $n$ vertices $V$ with an initial coloring
$c_0:V\to P$ using only the basic colors. The online input is a sequence of edges $e_1,e_2,\ldots$,
where each edge is an unordered pair of distinct vertices. We define
the graph $G_i$ by $G_i=(V,\{e_1,\ldots,e_i\})$. In response
to the arrival of each edge $e_i$, the algorithm outputs a coloring
$c_i:V\to P^*$ such that none of the edges in $G_i$  is
monochromatic. 

Given an algorithm $A$, an initial coloring $c_0$ and an edge sequence 
$\sigma=(e_1,\ldots,e_\ell)$, %
                              $A(c_0,\sigma)$
is a sequence of colorings $c_1,\ldots,c_\ell$.
Define the cost of $A$ on an instance
$(c_0,\sigma)$  as
$$
\Cost_A(c_0,\sigma)=
\sum_{i=1}^\ell\sum_{v\in  V}\left(1-\delta_{c_{i-1}(v)c_{i}(v)}\right)\Cost(c_i(v))~,
$$
where $\delta$ is
the Kronecker delta. That is, whenever a vertex is recolored, the
algorithm pays the cost of its new color.
In this paper we assume that the input is such that
the final graph can be colored using $P$, i.e., $G_\ell$ is $k$-colorable.
Given an initial coloring $c_0$, the \emph{offline cost} of an input sequence $\sigma$ with respect to the basic colors $P$ is %
$$
\Off_k(c_0,\sigma)=
\min\left\{\sum_{v\in  V}\left(1-\delta_{c_{0}(v)c^*(v)}\right)
\mid {\textrm{$c^* : c^*$ is a valid $k$-coloring of $G_i$ by $P$}} \right\}~.
$$
In other words, the offline cost of $\sigma$ w.r.t.\ $k$ is the least cost of
recoloring $V$  using basic colors so that $G_i$ has no monochromatic edges.

Let $I=(c_0,\sigma)$ denote an instance. 
The competitive ratio of an algorithm $A$ with respect to $k$ is
$\sup_I\left\{\Cost_A(I)/\Off_k(I)\right\}$. %
Let us make a quick observation about the problem.
\begin{lemma}
  \label{lem:cr=D}
  Any instance $I$ of recoloring with $k$ basic colors can be solved 
  by an online algorithm using
  $n-k$ special colors at cost at most $2D\cdot\Off_k(I)$, where $n$ is the
  number of vertices and $D$ is the cost of a special color.
\end{lemma}
\Proof 
Consider the final graph, with vertices colored by the initial
coloring. Any feasible solution must 
recolor a vertex cover of the monochromatic edges in this
graph. Therefore, denoting the size of such a minimum vertex cover by
$q$, we have that $\Off_k(I)\ge q$.  On
the other hand, we can maintain in an online manner a 2-approximate vertex cover
(e.g., using
an online version of the algorithm of Bar-Yehuda and
Even~\cite{LT_WVC}), and every time a vertex enters the online
cover, we recolor it
with a distinct new color. The special colors never collide 
because each is used at most once, and hence the total cost is at most $2qD$. 
\QED

\paragraph{Additional notation.}
\begin{itemize}
\item Given a sequence $\sigma$, $\sigma[i]$  denotes the
  prefix of the first $i$ elements of $\sigma$. Given sequences
  $\sigma$ and $\sigma'$,  $\sigma\circ\sigma'$ denotes the
  sequence obtained by concatenating $\sigma$ and $\sigma'$.
\item Given a graph $G=(V,E)$ and $V'\subseteq V$, 
   $G[V']$  denotes the graph induced by $V'$, i.e.,
  the graph with vertices $V'$ whose  edges are all edges of $E$ with
  both endpoints in $V'$.
\end{itemize}

\paragraph{Graph bond.}
The largest bond size is a graph parameter related to max cut, but it is quite different. Intuitively,
the graph bond is a measure of how close the graph is to a tree. We give below a
definition that generalizes the standard one (e.g., \cite{bond}) to
graphs with multiple connected components.
\begin{definition}
\label{def:bond}
    Let $G=(V,E)$ be a graph with $k\ge1$ connected components. An edge set $B\subseteq E$ is a \textbf{bond} of $G$ if $(V,E\setminus B$) has exactly $k+1$ connected components. The size of the largest bond of $G$ is    
    denoted $\beta(G)$.
\end{definition}
Note that $\beta(G)=1$ if and only if $G$ is a forest. Also note
that the size of the largest bond of a graph is monotone in its edge set.
More formally, if $G=(V,E)$ and $ G'=(V,E')$ are two graphs with the same vertex set, then  $E \subseteq E'$ implies $\beta(G)\le\beta(G')$.

We remark that an alternative definition for bond is a \textit{minimal edge cut}
\cite{10.5555/1481153}: an edge cut 
of $G=(V,E)$ is an edge set $B\subseteq E$ such that $(V,E\setminus B)$ has more connected components than $G$, and an edge cut $B$ is \textit{minimal} if no proper subset of $ B$ is an edge cut of $G$.

\section{Non-Uniform Cost}
\label{sec:non-uniform}
In this section we prove our main result. %
Intuitively, we show that while the competitive ratio for general bipartite graphs is known to be $\Theta(\log n)$,
by using $\Delta+1$ additional colors of cost (at most) $D$, one can reduce the competitive ratio
to $\Theta(\log D)$ in graphs whose largest bond is small. In 
\ifarxiv
\sectionref{sec:A-non-uniform}, 
\else
the full version,
\fi
we show that we can achieve the same competitive ratio with only $\Delta$ additional colors, using a somewhat more complicated algorithm.

Our algorithm uses 
the algorithm of Azar \textit{et al.}\ \cite{AMPT-22} for bipartite graphs as a black box;
henceforth, we refer to Algorithm 1 of \cite{AMPT-22} as $\alg$.
The basic strategy of our algorithm is as follows. The input sequence
of edges $\sigma$ is split in two: a subsequence denoted
$\sigma^{\rm{sim}}$ is sent to a simulation by algorithm \Alg (which uses
only the two basic colors), and the remaining edges
are sent to a procedure
called $\recx$, which 
uses the additional special colors. Determining which edge
goes to $\sigma^{\rm sim}$ depends on the size of the connected components of its two endpoints,  and by the number of vertices already recolored by \Alg in
them. The idea is to control the simulation of $\alg$ so that its cost remains
within the range of $O(\log D)$ competitiveness, and use the special colors 
only once we know that we can pay for them. The bond size affects the total 
cost due to the edges that are not sent to the simulation.

\subsection{Algorithm $\cB$: Specification}
To describe the algorithm, we need some notation.
Recall that $\sigma[i]$ is the sequence of the first $i$
edges of $\sigma$. 
We use $E_i$ to denote the set of edges in $\sigma[i]$:
$E_i=\Set{e_1,e_2,\ldots e_i}$. We use $\sigma^{\rm{sim}}_i$ to denote the
subsequence of edges sent to the simulation of \Alg in steps
$1,\ldots,i$, and $E[i]_{sim}$ to denote the corresponding set of edges. $R_i$ is used to
denote the set of vertices ever recolored by algorithm \Alg when
executed on input $\sigma^{\rm{sim}}_i$. We also define $R$ to be $R_{|\sigma|}$.  

We now specify the algorithm, using the parameters $D$ (the maximal cost of
a color), and $\alpha\in(0,1)$, a  parameter of our choice which indirectly controls how many of the edges will be
sent to the simulation: the smaller $\alpha$ is, less edges will be sent to the simulation.

We also  define the following concepts.
\begin{definition}
\label{def:moderate}
  Consider a sequence of edges $\sigma$ and a connected component $C$ of the graph which is edge-induced by the set of edges in $\sigma$.
  \begin{itemize}
  \item $C$ is  \textbf{small} if $|C|\le D$ and \textbf{large}
    otherwise, where $|C|$ is the number of vertices in $C$.
  \item $C$ is  $\sigma$-\textbf{light} if $|R\cap C|/|C| \leq \alpha$,
    and 
    $\sigma$-\textbf{heavy} otherwise, where $R$ is the set of vertices ever recolored by algorithm \Alg on input $\sigma$.
   
  \end{itemize}
\end{definition}

\begin{definition}
\label{def:sequences}
Consider an input sequence $\sigma$, and two values $D$ and $\alpha$.
  \begin{itemize}
  \item An input sequence $\sigma$ is called \textbf{$(D,\alpha)$-moderate}
  if for every edge $e_i=(u,v)$ in $\sigma$, it holds that if $u$ and $v$ are in two distinct connected components $C_u$ and $C_v$ with respect to $e_1,\ldots,e_{i-1}$, then either $C_u$ or $C_v$  is either small or $\sigma[i-1]$-light (i.e., no edge in $\sigma$ connects two distinct components which are large and $\sigma[i-1]$-heavy).
  
  \item  The two series of subsequences $\sigma^{\rm sim}_{i}$ and $\sigma^{{\rm exc}}_i$ of $\sigma$ are defined inductively as follows.
\begin{align*}
    \sigma^{{\rm sim}}_i&=\begin{cases}
    \bot&\text{if $i=0$}\\
    \sigma^{{\rm sim}}_{i-1} \circ e_i&
    \text{if $i>0$ and $\sigma^{{\rm sim}}_{i-1} \circ e_i$ is $(D,\alpha)$-moderate}\\
    \sigma^{{\rm sim}}_{i-1}&\text{if $i>0$ and $\sigma^{{\rm sim}}_{i-1} \circ e_i$ is not $(D,\alpha)$-moderate}      
\end{cases}
\\
\sigma^{{\rm exc}}_i&=\begin{cases}
    \bot&\text{if $i=0$}\\
    \sigma^{{\rm exc}}_{i-1}&
    \text{if $i>0$ and $\sigma^{{\rm sim}}_{i-1} \circ e_i$ is $(D,\alpha)$-moderate}\\
    \sigma^{{\rm exc}}_{i-1} \circ e_i&\text{if $i>0$ and $\sigma^{{\rm sim}}_{i-1} \circ e_i$ is not $(D,\alpha)$-moderate}      
\end{cases}
\end{align*}

\end{itemize}
\end{definition}
(Informally, $e_i$ is appended to $\sigma^{{\rm sim}}_{i-1}$  if it keeps the moderation property, and to $\sigma^{{\rm exc}}_{i-1}$ otherwise.)
The pseudocode for the algorithm appears in the following page.

\subsection{Analysis}

We now turn to analyze the algorithm. %

The interesting part about the algorithm is that the simulation of $\alg$ is
unaware of some of the edges. 
We start by  showing that Algorithm $\cB$ produces a valid coloring.  

\begin{lemma}
  \label{lem:valid}
  After every step $i$, $c_i$ is a valid coloring of the graph $G_i=(V,E_i)$.
\end{lemma}
\Proof
First observe that once a vertex becomes special, then it is always colored by a special color.   Likewise, only special vertices are colored by special colors.

\noindent
\hspace{-1mm}\fbox{\vbox{\small
\vspace{1mm}
\noindent
\textbf{Algorithm $\cB$}
\vspace{2mm}\\
\textbf{State:}
\begin{itemize}\vspace{-1mm}
\item Each vertex $u$ has an \emph{actual} color $c(u)$
  and a \emph{simulated} color
  $\C(u)$. The simulated coloring $\C$ is the coloring maintained by the simulation of $\alg$. The initial actual colors are given as input, and the
  initial simulated colors are the initial actual colors.
\item Each vertex records whether its simulated color was ever changed by 
the simulation of $\alg$. 
This
allows the algorithm to maintain the set $R_i$.
\item Each vertex has an indication whether it is ``special" or not. Initially all vertices are not special.
\end{itemize}

\noindent
\textbf{Action:}\\
Upon the arrival of edge $e_i = (u_i, v_i)$:%
\begin{enumerate}[\hspace{1em}a.]
\item 
If $\sigma^{\rm{sim}}_{i-1}\circ e_i$  is
    $(D,\alpha)$-moderate then \\
    \hspace*{5mm} send $e_i$ to \Alg   (which updates $\C(\cdot)$);
    \hfill \textit{//$\sigma^{\rm{sim}}_{i}= \sigma^{\rm{sim}}_{i-1}\circ e_i$ } \\
    \hspace*{5mm} set $c(w):=\C(w)$ for  every non-special vertex $w$.
\item Else \\
\hspace*{5mm}If both $u_i$ and $v_i$ are not special then mark $u_i$ as {\em special}.
\item   
\label {st:special}
\hspace*{4mm} Invoke $\recx(u_i,v_i)$
\label{st:simulate}
\end{enumerate}

\noindent
\textbf{Procedure} $\recx(u,v)$:%
\begin{enumerate}%
\item If $u$  and $v$ are special  then \\
\hspace*{5mm} if $e=(u,v)$ is monochromatic then \\
\hspace*{10mm}  recolor $u$ with a free special color. \hfill\textit{// there are $\Delta +1$ special colors}
\item  Else \\
\hspace*{5mm}   If $u$ is special then \\
\hspace*{10mm}  if $u$  is colored by a basic color then   \\
\hspace*{15mm} recolor $u$ with a free special color. 
\hfill\textit{// first time $u$ is colored special}
\item \hspace*{5mm}  Else \\
 \hspace*{10mm}  If $v$ is special then \hfill\textit{// this case cannot happen because of the way recx is used.}\\
\hspace*{15mm}  if $v$  is colored by a basic color then   \\
\hspace*{20mm} recolor $v$ with a free special color. 

\end{enumerate}
}
}
\\

We
prove the lemma by induction on $i$. The base case is $i=0$, in which $E_0=\emptyset$ and
hence any coloring is valid. For the inductive step, we first consider all edges but the new edge $e_i$ and for those we proceed in two 
sub-steps. Then we consider the new edge $e_i$.

For all edges but $e_i$,
if $e_i$ is not a simulated edge and is not sent to $\cA$, then the first 
sub-step is empty.
If $e_i$ is  a simulated edge and is sent to $\cA$, then the colors of some non-special vertices may change.
After this sub-step, for each (old) edge: (1)  if  its two endpoints are non-special then the edge is not monochromatic by the correctness of $\cA$; (2) if its two endpoints are special,  then  their color does not change in this sub-step and the edge is not monochromatic by the induction hypothesis; (3) if one endpoint is special and the other is not, then one and only one endpoint is colored by a special color, and hence the edge is not monochromatic. 

The second sub-step is the invocation of $\recx$ on the input edge. Following this invocation one  vertex might be recolored to a special color. Clearly any (old) edge with at least one node not special remains non-monochromatic by the induction hypothesis. For an (old) edge with two special endpoints, if their color is not changed, they remain non-monochromatic by the induction hypothesis. For such an edge for which  one of its endpoints changed color by $\recx$, the code %
ensures that it is non-monochromatic after the execution of $\recx$. 

Now as to the edge $e_i$ itself, we have a number of cases depending whether it is a simulated edge and the status (special or not) of its two endpoints when step $i$ begins.
\begin{enumerate}[1.]
\item If its two endpoints are non-special when step $i$ starts:
\begin{itemize}
\item if $e_i$ is a simulated edge: at the end of the first sub-step $e_i$ is not monochromatic by the correctness of $\cA$; none of its endpoints is marked special and hence $\recx$ does not recolor any node and $e_i$ remains non-monochromatic.
\item  if $e_i$ is not a simulated edge: One of its endpoints is marked special; $\recx$ recolors that vertex by a special color, while the other endpoint remains colored by a basic color; hence $e_i$ is not monochromatic.
\end{itemize}
\item  If its two endpoints are special when step $i$ starts, then  regardless of whether $e_i$ is a simulated edge or not the code of $\recx$ ensures that $e_i$ is not monochromatic.
\item \vspace*{-3mm}If one endpoint is special and the other is not  when step $i$ starts, then regardless  of whether $e_i$ is a simulated edge or not, the status of neither vertex changes, and one of them remains colored by a special color and the other by a basic color, hence $e_i$ is not monochromatic.
\QED
\end{enumerate}

\vspace*{-2mm}
We now turn to analyze the competitiveness of our algorithm. To this end, 
we first prove the following property  about
graphs with bounded bond size.

\begin{lemma}
  \label{lem:bond}
  Let $G=(V,E)$ be a connected graph  with largest bond size at most $\beta$, and let
  $\Set{V_1,\ldots,V_k}$ be a partition of $V$ such that the induced
  subgraph $G[V_i]$ is connected, for every $1\le i\le k$. Then the
  number of edges which are not contained in any of these induced subgraphs
  (i.e., the number of edges with endpoints in two different  parts of the
  partition) is at most $(k-1)\beta$.
\end{lemma}
\Proof
Consider the multi-graph $G'=(V',E')$ obtained from $G$ by contracting each $V_i$ to a single node and eliminating self-loops. 
We shall prove that  $|E'| \leq  (k-1)\beta$. 

First, we note that $|V'|=k$, and that the size of the largest bond of  $G'$ is at most $\beta$.
We now prove, by induction on $k$, that if a 
loop-free multigraph 
$G'=(V',E')$ has $k$ nodes and bond size at most $\beta$, then $|E'| \leq  (k-1)\beta$.
The base case is $k=1$; in this case,  $E'$ is empty 
since $G'$ does not contain self loops.

For the inductive step, fix any $k\ge2$. 
Let $v$ be an arbitrary node in $V'$, and let $\bar{V}=V' \setminus \Set v$.
We proceed according to the connectivity of $\bar{V}$.
If $G'[\bar{V}]$ is connected, then the degree of $v$ is at most $\beta$. Since $|\bar{V}| = k-1$, by the induction hypothesis and the fact that the size of
the largest bond of $G'[\bar{V}]$ is at most the size of the largest bond of $G'$ (which is at most $\beta$), we have that $|E'| \leq \beta +(k-2)\beta = (k-1)\beta$.

Otherwise, $G'[\bar{V}]$ is disconnected. Let ${C}_1$ be the set of vertices of an arbitrary  %
connected component  of $\bar{V}$,  
and denote ${C}_2=\bar{V} \setminus  C_1$.  
Let $C_1^v= {C}_1 \cup \{v\}$,  and $C_2^v= {C}_2 \cup \{v\}$. 
Trivially $|C_1^v|+|C_2^v|=k+1$, and $m_1+m_2=|E'|$,  where $m_1$ and $m_2$ are the number of edges in $G'[C_1^v]$ and  in $G'[C_2^v]$, respectively. %
Since the largest bond of a subgraph is no larger than  the largest bond of the original graph, 
and since the number of nodes in each of $G'[C_1^v]$ and $G'[C_2^v]$ is strictly less 
than $k$, by the inductive hypothesis  we have that 
$m_1 \leq (|C_1^v| - 1)\beta$ and $m_2 \leq (|C_2^v| - 1) \beta$.
It follows that $|E'|=m_1 + m_2 \leq (k-1)\beta$.
\QED

\begin{corollary}
  \label{cor:bond}
  Let $G=(V,E)$ be a  graph with $\beta(G)\le\beta$, and let
  $C_1,\ldots,C_k$ be a collection of disjoint  subsets of vertices in $V$ such that 
  $G[C_i]$ is connected for all $1\le i\le k$. Then 
  $
  \left|\Set{(v,u)\in E\,:\, v\in C_i, u\in C_j, i\ne j}\right|\le\beta(k-1)~,
  $
  i.e., 
  the number of edges with endpoints in two different subsets
  is at most $\beta(k-1)$.
\end{corollary}
\Proof
Let $V_1,\ldots,V_k,V_{k+1}$ be any partition of $V$
such that:
\begin{itemize}
    \item  for each $1\le i\le k$, $C_i\subseteq V_i$, and $G[C_i]$ is connected; and
    \item $V_{k+1}$ is exactly the set of vertices which are not connected in $G$ to any vertex of $\bigcup_{i=1}^kC_i$.
\end{itemize}
This can be achieved by associating each vertex of $V\setminus V_{k+1}$ with the set $C_i$ closest to it (like in a Voronoi partition). 
Note that we may assume without loss of generality that $V_{k+1}$ is empty:
otherwise, consider the graph $G'\DEF G[V\setminus V_{k+1}]$:
Clearly, by the definition of $V_{k+1}$,
the number of  edges connecting vertices in different $C_i$
in $G$ and $G'$ is the same.

Since every edge between two subsets $C_i,C_j$ connects different parts of the partition $V_i$ and $V_j$, 
the result %
follows from \lemmaref{lem:bond}, when applied to each connected component of $G$ separately.
\QED

We proceed to bound from above the cost of Algorithm $\cB$. 
The cost consists of two
parts: the cost incurred by Procedure $\recx$ and the cost due to the
simulation of \Alg. 
We start by stating a number of facts due to the definition of the algorithm.
\begin{fact}
\label{fa:special_color_big_comp}
A vertex which is colored by a special color after step $i$ must belong to a connected component of $E[i-1]_{{\rm sim}}$ that has more than $\alpha D$ vertices of $R$.
\end{fact}
\Proof
Observe that a node $v$ is colored by a special color only in procedure $\recx$,  when it is a special vertex. Moreover, it is marked special only when an edge $e_j=(v,u)$ arrives, and that edge is in $E \setminus E_{{\rm sim}}$. 
It follows that there is an edge $e_j$, for $j \leq i$, which  connects two large and heavy connected components of $E[j-1]_{{\rm sim}}$. Hence,  $v$ is part of a large connected component of 
$E[j-1]_{{\rm sim}}$ which has more than $\alpha D$ vertices of $R$.
\QED

Recalling the sequences  $\sigma^{{\rm exc}}$ and $\sigma^{{\rm sim}}$ defined in \defref{def:sequences} as a function of any sequence $\sigma$,
 we prove the following lemma that allows us to (indirectly) give an upper bound on the number of vertices that require ``special treatment". %

\begin{lemma}
\label{le:special_edges}
For any given  $\sigma$, $D$ and $\alpha$, it holds that $|\sigma^{{\rm exc}}| \leq \frac{\beta|R|}{\alpha D}$, where $R$ is the set of vertices whose colors are modified when $\sigma^{{\rm sim}}$  is given as input to \Alg.
\end{lemma}

\Proof
In order to prove the lemma we maintain, as the input sequence $\sigma^{{\rm sim}}$ proceeds, sets of vertices, $B_1,B_2,\ldots,B_s$, which are pairwise disjoint; furthermore, all vertices in each $B_i$ belong to the same connected component of the input graph (with respect to all edges, not only those in $\sigma^{{\rm sim}}$). We call the $B_i$ sets {\em witness sets}.
We construct the witness sets online as follows.
Initially, %
there are no witness sets.
When an edge  $e=(u,v) \in \sigma $ arrives, we modify the witness sets as follows. 
\begin{enumerate}[I]
\item If none of $\Set{u,v}$ is already in one of the existing witness sets, and if the connected component that
  contains $e$ (after $e$ is added to the graph)  has at most $\alpha D$ vertices from $R$: do nothing.
\item 
\label{action2}
If none of $\Set{u,v}$ is already in one of the existing witness sets, and if the connected component that
  contains $e$ (after $e$ is added to the graph) has more than $\alpha D$ vertices from $R$: create a
  witness set %
  which contains all the vertices in that connected component.
\item 
\label{action3}
If one of $\Set{u,v}$,  w.l.o.g.  $u$,  is already in an existing witness set, say $B$, and  $v$ is not: add all the vertices of the connected component of $v$ to $B$. (As we prove below in Point~\ref{pt:does_not_change}, these vertices do not yet belong to any witness set.) 
\item If both $\Set{u,v}$ are already in some witness sets (possibly the same one): do nothing. 
\end{enumerate}

Intuitively, witness sets are created when a
connected component passes the
``critical mass'' threshold of $\alpha D$ vertices from $R$, and
vertices that do not belong to any witness set  are added to the witness set they encounter, i.e., the set of the first 
vertex that belongs to a witness set, to which they are connected.

The following properties follow from a straightforward induction on the input sequence.
\begin{enumerate}
\item
\label{pt:same_component} All the vertices of a given witness set  are in the same connected component of $\sigma$.
This is because a new witness set is created  from connected vertices. Moreover, additions to a witness set are always in the form of vertices connected to vertices already in that witness set.
\item 
\label{pt:in_sets} All the vertices in a connected component of $\sigma^{{\rm sim}}$ with more
  than $\alpha D$ vertices of $R$ are in some  witness set (not necessarily all in the same witness set). This follows by induction on the arrival of new edges of $\sigma^{{\rm sim}}$ and the actions relative to witness sets taken when this happens.

\item \label{pt:both_in_or_out}
Every two vertices $u,v$ which are in the same connected component of $\sigma$ are either both in some witness set (not necessarily the same), or both are not in any witness set. This follows from the fact that the actions taken regarding witness sets exclude the possibility of an edge with exactly one endpoint in a witness set.
  \item
  \label {pt:does_not_change} 
  If a vertex $w$  belongs to some witness set $B$ at some
  point of the execution, then $w$ belongs to $B$ in the remainder of the execution.
 This follows from the fact that a vertex $w$  becomes 
 a member of witness set $B'$ only in Cases~\ref{action2} or~\ref{action3} above. 
 
 In Case~\ref{action2}, $w$ did not belong before to any witness set because both $u$ and $v$ did not belong to any witness set and $w$ is in the same connected component of either $u$ or $v$. Hence, by Point~\ref{pt:both_in_or_out}, $w$ is not in any witness set.

 In Case~\ref{action3}, similarily, $w$ is connected to $v$ and since $v$ does not belong to any witness set, then by Point~\ref{pt:both_in_or_out} $w$ is not in any witness set.
\end{enumerate}

   It follows from the above properties that 
the total number of witness sets is at most $\frac{|R|}{\alpha
  D}$, because each witness set that is created requires at least $\alpha D$ distinct vertices of $R$, vertices do not change their witness set, and the sets are pairwise disjoint.

Next, we claim that when an edge $e$  is added to $\sigma^{{\rm exc}}$ (i.e., it connects two distinct large-and-heavy connected components 
$C,C'$ of $\sigma^{{\rm sim}}$),
then the vertices of $C$ and $C'$ already belong to two distinct witness sets.
By definition, each of $C$ and $C'$ contains more than $\alpha D$
vertices of $R$, and hence, by Point~(\ref{pt:in_sets}) above, the endpoints of $e$
are already associated with witness sets. To show that the endpoints of $e$  belong to distinct witness sets, assume,
by way of contradiction, that the endpoints of $e$ belong
to the same witness set. Then from
Point~\ref{pt:same_component} we have that $C$ and $C'$ are already connected in $\sigma$ (but, by assumption, they are not connected in $\sigma^{{\rm sim}}$). 
Let $e_0$ be the first edge (according to the order of the arrival of the edges)  in $\sigma^{{\rm exc}}$  that completes a path from some node in $C$ to some node in 
$C'$. Since $e_0\notin \sigma^{{\rm sim}}$, it must be the case that when $e_0$ arrived, 
both of its endpoints belonged to two large and heavy connected components of $\sigma^{{\rm sim}}$. Point~\ref{pt:in_sets}  ensures then that the vertices of $C$ and $C'$ are already in  witness sets.  
Point~\ref{pt:same_component} implies that these witness sets must be distinct, 
because before the arrival of $e_0$, $C$ and $C'$ are not connected to each other
in $\sigma$.  The sets to which the vertices of $C$ and $C'$ belong are the same, and hence distinct, when $e$ arrives, by Point~\ref{pt:does_not_change}.
This is in contradiction to the assumption that the witness
sets are the same.

Thus, an edge is added to $\sigma^{{\rm exc}}$ only when its endpoints are in two \emph{distinct} witness sets.
Since there can be  at most $\frac{|R|}{\alpha D}$ such sets,
we get from \corollaryref{cor:bond} and the assumption that the graph has bond size
at most $\beta$, that there are at most
$\frac{\beta|R|}{\alpha D}$ such edges.
\QED

\begin{lemma}
  \label{lem:special edges}
  The total cost incurred in procedure $\recx$ is 
  $O(\beta^2|R|/\alpha)$, where $R$ is the set of vertices ever
  recolored by \Alg.
\end{lemma}
\Proof
Procedure  $\recx$ only recolors vertices that are marked special. For each special vertex $v$, the procedure $\recx$ recolors $v$ from a basic color to a special color exactly once in the step when $v$ becomes special, and might recolor $v$ again when a new edge $(v,u)$ arrives, and also $u$ is marked special.
Let $V'$ be the set of vertices ever marked special, and $E'$ the set of edges such that when they arrive both their endpoints are marked special. 

A node $v$  is marked special only when an edge $e_i=(v,u)$ arrives and $e_i \in  E \setminus E_{{\rm sim}}$. By Lemma~\ref{le:special_edges} $| E \setminus E_{{\rm sim}}| \leq O(\frac{\beta|R|}{\alpha D})$, and hence $|V'| \leq O(\frac{\beta|R|}{\alpha D})$.

Let $|V'|=\ell$, and   $V' = \{ u_j : 1 \leq j \leq \ell\}$.
By applying
\corollaryref{cor:bond} with sets $C_1 = \{u_1\}, \ldots, C_\ell = \{u_\ell\}$, we have that 
$|E'| \leq O(\frac{\beta^2|R|}{\alpha D})$.

The cost of $\recx$ is therefore $O(|V'| + |E'|)\cdot D =  O(\frac{\beta|R|}{\alpha D} + \frac{\beta^2|R|}{\alpha D}) \cdot D = O(\frac{\beta^2|R|}{\alpha}).$
\QED

Next, we bound from above the cost incurred by the simulation of \Alg.
We need a refined analysis of Algorithm $\alg$.
In the following two lemmas, we recall  properties of \Alg %
from~\cite{AMPT-22}.
The first lemma relates the number of vertices recolored by \Alg to
the optimal cost.
\begin{lemma}[\cite{AMPT-22}, Lemma~3.4]
  \label{lem:size}
  Let $\Opt_2[i]$ denote the offline cost for input $\sigma[i]$ when
  using two colors of unit cost. Then
  $|R_i|\le 3\cdot\Opt_2[i]$.
\end{lemma}
The next lemma is used to upper-bound the number of times a vertex is
recolored by \Alg. %
\begin{lemma}[\cite{AMPT-22}, Proposition 3.5]
    \label{lem:I+}
    \renewcommand{\Cost}{\mathbf{r}}
    Let $\Cost(i)$ denote the number of vertices recolored by
    \Alg in step $i$. There exists a subset of steps
    $I^+\subseteq\Range{|\sigma|}$ such that
    $\sum_{i \in I^+}\Cost(i)\ge \frac17\sum_{i=1}^{|\sigma|}\Cost(i)$.
    Moreover, for every $i \in I^+$, if the arriving
    edge connects two distinct connected components $C$ and $C'$, and \Alg recolors $C$, then
    $|C'\cup C| \geq \frac{5}{4}|C|$.
  \end{lemma}
  
In \cite{AMPT-22}, \lemmaref{lem:size} and \lemmaref{lem:I+} are used
to show an $O(\log n)$ bound on the competitive ratio of $\alg$. Here, we prove a tighter bound on the competitive ratio 
when the input sequence is moderate (cf.~\defref{def:moderate}).
The proof uses amortized analysis by designing an appropriate charging scheme.

\begin{lemma}
  \label{lem:moderate}
  If $\sigma$ is a $(D,\alpha)$-moderate input sequence,
  then $\Cost_\alg(\sigma) \leq O(|R|\log D/ (1-\alpha))$.
\end{lemma}

\Proof
Let $R_i$ be the set of vertices
recolored by \Alg so far after it processed the $i$th input edge. Let $I^+$ be the
subset of steps given by Lemma~\ref{lem:I+}; it suffices to bound from above the
cost incurred by \Alg on steps in $I^+$. We will do this via a
charging
scheme. %

The high-level intuition behind the charging scheme is as follows. Let
$i \in I^+$, and suppose $e_i$ connects components $C_i$ and $C'_i$,
and that in response, \Alg recolors $C_i$ (recall that we consider bipartite graphs and two colors, hence a monochromatic edge must connect two distinct connected components). Our goal is to charge a the load of $1$ on 
certain vertices and show that (1) the total charge in each step $i$ is at least a 
constant fraction of $|C_i \cap R|$ or $|C'_i \cap R|$ and (2)  every vertex is charged $O(\log D)$ times in total over the course of the input sequence. There are {\em four} cases to consider and each case will use a different type of charging. (1) The first is when $C_i$ is small,
i.e.~$|C_i| \leq D$. In this case, we  charge every vertex in $C_i$.  Lemma~\ref{lem:I+}   implies that a vertex can only be charged by this charging type a total of $O(\log D)$ times. 
(2) The second case is when $C_i$ is large and light. In this case, we charge to the vertices of $C_i$ that are newly recolored by the algorithm. Since $C_i$ is light, the number of vertices that are recolored for the first
time by \Alg in this step is sufficiently large. Since this type of charging charges newly recolored vertices, a vertex can be charged by this charging type at most once.

The remaining possibilities concern case (3) when $C_i$ is large and heavy. We  consider two subcases. (3.1) The first subcase is when $C_i$ is large and heavy and $C'_i$ is small and heavy. In this case, we charge to $C'_i \cap R$. Lemma~\ref{lem:I+} implies that $|R \cap C'_i|$ is at least an $\alpha$  fraction of $|C_i|$. Moreover, since the vertices in $C'_i$ are part of a large component after this step, a vertex can be charged by this charging type at most once. The final and most interesting case is (3.2) when $C'_i$ is light. To handle
this case, we maintain, for each (maximal) connected component $C$ that exists at a given time, a subset $X(C)$ of vertices. We will charge to the vertices of $X(C_i)$ in this case. The sets $X$ are defined inductively as 
described in the pseudocode of the charging scheme below. As we prove below, each vertex is charged by this type of charging only once, we the size of the sets $X$ is large enough to compensate for the cost of recoloring.

Note that these cases are exhaustive as $C_i$ and $C'_i$ cannot be both large and
heavy, otherwise the edge would not have been input to the simulation $A$. The following summarizes in a formal manner the charging scheme described above. We then formally prove the desired properties of the charging scheme as stated informally above.

\noindent\hspace*{-1mm}\fbox{\vbox{\small
\noindent
\textbf{Charging scheme}%

\begin{itemize}

\item
Initially, $X(\{v\}) = \{v\}$ for every vertex $v$.
\item 
When a new edge $e_i = (u_i,v_i)$ arrives connecting components $C_i$ and $C'_i$:
\begin{enumerate}
\item Suppose \Alg recolors $C_i$
  \begin{enumerate}
  \item Charge vertices  as follows: %
    \begin{enumerate}
    \item\label{small} If $C_i$ is small, then charge $1$ to each vertex in $C_i$
    \item\label{large-1} If $C_i$ is large and light, then charge $1$
      to each vertex in $C_i \setminus R_{i-1}$, i.e., the vertices of
      $C_i$ that were never recolored before step $i$.
    \item\label{large-2} If $C_i$ is large and heavy, $C_i'$ small and heavy, then charge $1$ to each vertex in $R_{i-1}(C_i')$
    \item\label{large-3} If $C_i$ is large and heavy, $C_i'$ light, then charge $1$ to each vertex in $X(C_i)$.
    \end{enumerate}
    \item \label{update-x} If $C_i'$ is light, then $X(C_i \cup C_i') = (C_i \cup C_i')  \setminus R_i$; else $X(C_i \cup C_i') = X(C_i) \cup X(C_i')$  
    \end{enumerate}
    \item If \Alg does not recolor, then $X(C_i \cup C_i') = X(C_i) \cup X(C_i')$
\end{enumerate}
\end{itemize}
}
}

Observe that every vertex charged is either in $C_i$,  the component
being recolored, or was previously recolored (case
\ref{large-2}). Thus every vertex that is charged is in $R$. It now
remains to show that in each step, the number of vertices being charged is
at least a constant fraction of the number of vertices being
recolored---i.e.~the total charge received by the vertices can pay for
the recoloring cost---and that each vertex is not charged more than
$O(\log D)$ times.

\begin{claim}
  \label{claim:X}
  Consider a connected component $C$ that exists after an arbitrary step $i$, and the set $X(C)$ defined by the charging scheme. For every $C \in \cP$, we have  $|X(C)| \geq |C|(1-\alpha)/5$.
\end{claim}

\Proof We prove this claim by induction on $|C|$, the base case being $|C|=1$. By definition for $C$, such that $|C|=1$, $X(C)=C$ and hence the claim hold, since $0 < \alpha <1$.  For the inductive step, suppose that $C = A \cup A'$,
and that $|X(A)| \geq |A|(1-\alpha)/5$ and
$|X(A')| \geq |A'|(1-\alpha)/5$. 
If $X(C) = X(A) \cup X(A')$ (which occurs in two distinct cases in the specifications of the charging scheme)
then, since  $A \cap A'=\emptyset$,
$$
|X(C)| = |X(A)| + |X(A')| \geq (|A| + |A'|)(1-\alpha)/5 =
|C|(1-\alpha)/5.
$$
Otherwise, 
$X(C) = (A \cup A') \setminus R_i$.  Then,
when $e_i$ arrived, \Alg recolored one of the two connected components $A, A'$ and the other connected component was
light. Suppose w.l.o.g. that  \Alg recolored $A$, and $A'$ was light. Then, we have
$$
C \setminus R_i \subseteq A' \setminus R_i = A' \setminus
R_{i-1}~.$$
By the lightness of $A'$, we have that
$|A' \setminus R_{i-1}| \geq |A'| (1-\alpha) \geq |C|(1-\alpha)/5$ by
Lemma~\ref{lem:I+}. This concludes the induction argument. 
\QED
\begin{claim}
  \label{claim:charge-1}
  Suppose \Alg recolors component $C_i$ at step $i$. Then the number
  of vertices being charged is at least   $\Omega(|C_i|(1-\alpha))$.
\end{claim}
\Proof
We show that the number of vertices being charged is at
least $|C|(1 - \alpha)/20$ in every case. In case \ref{small}, we
charge $C$ so this clearly holds. In case \ref{large-1}, lightness of
$C$ implies that
$|C \setminus R_{i-1}| \geq |C|(1-\alpha)/5 \geq |C|(1-\alpha)/20$. In
case \ref{large-2}, we have
$|R_{i-1}(C')| \geq |C'|(1-\alpha)/5 \geq |C|(1-\alpha)/20$ where the
first inequality is due to $C'$ being heavy and the second due to
Lemma~\ref{lem:I+}. For case \ref{large-3}, we get that
$|X(C)| \geq |C|(1-\alpha)/5$ from Claim~\ref{claim:X}.  \QED

\begin{claim}
  \label{claim:charge-2}
  Every vertex is charged at most $O(\log D)$ times.
\end{claim}
\Proof
By \lemmaref{lem:I+}, the number of times a vertex $v$ is
charged due to case \ref{small} is $O(\log D)$. The number of
times a vertex $v$ is charged due to line \ref{large-1} is at most
once since it is recolored and now belongs to $R_i$. The number of
times it can be charged due to \ref{large-2} is at most once since it
now belongs to a large component. The number of times it can be
charged due to \ref{large-3} is at most once since afterwards it
belongs to $R_i$ and thus cannot be in $X(A)$ for any future component, due to the way the set $X$ is modified in~\ref{update-x}.
\QED

Combining the
fact that only vertices in $R$ can be charged  and
Claims~\ref{claim:charge-1} and \ref{claim:charge-2},
we are done proving \lemmaref{lem:moderate}.
\QED

We are now ready to conclude the analysis of Algorithm $\cB$.
\ProofOf{Theorem}{thm:cost-D}
By
\lemmaref{lem:special edges}, the total cost incurred by recoloring using a
special color is $O(|R|\cdot\frac{\beta^2}{\alpha})$. 
Next, note that by the code the sequence given as input to $\alg$, 
$\sigma^{\rm{sim}}$, is $(D,\alpha)$-moderate. 
It therefore follows from  \lemmaref{lem:moderate} that the
total cost due to the simulation of \Alg is $O(|R|\log
D/ (1-\alpha))$. Since by \lemmaref{lem:size} the optimal cost is $\Omega(|R|)$,
we are done by picking, say, $\alpha=1/2$.
\QED

\section{Uniform cost}
\label{sec:uniform}

In this section, we still consider vertex recoloring of bipartite
graphs, but now in the setting where that all colors, basic and special, have the same cost.
We thus present an algorithm which has better competitive ratio if more than two
colors are available, covering the spectrum between an $O(\log n)$ competitive
ratio with no special colors, and a $O(\log\beta)$ competitive ratio with $O(\log_\beta n)$ 
special colors, where $\beta$ is an upper bound on the size of graph bonds.
Specifically, we prove \theoremref{thm:tradeoff}, reproduced below.

\tradeoffthm*

The idea is to use $O(\epsilon\log n)$ instances of
Algorithm $\alg$ (Alg.~1 of~\cite{AMPT-22}) for recoloring bipartite graphs.

\paragraph{High-Level Description.} 
We push the ideas of simulation and promoting vertices to be ``special'', used in algorithm $\cB$, even further. Roughly speaking, the main idea of the algorithm is to create multiple (possibly overlapping) subinstances of the problem, such that the input sequence of each subinstance is moderate, as in \defref{def:moderate}. Each subsequence is fed into
a different instance of $\alg$, which uses a distinct pair of colors.
At every point in time, each vertex is associated with
some instance $j$, which determines its actual color.

More specifically, denote the $j$th instance of $\alg$
by $\alg_j$, its input sequence by
$\sigma^j$, and the coloring that it maintains at any time  by $c^j$. Each vertex $v$ is associated at any given time with one of the instances, denoted $\Part(v)$. Initially, $\Part(v)=1$ for all $v\in V$.
We define  $V^j\DEF\Set{v\in V\mid\Part(v)=j}$.
$\alg_j$ determines the coloring of $V^j$: $c^j:V^j\to\Set{2j-1,2j}$. Our algorithm maintains the following invariants: (1) each subsequence $\sigma^j$ is moderate; (2) for each level $j$, the coloring $c^j$ is a valid coloring of $V^j$ with respect to $G[V^j]$ where $G$ is the graph consisting of every edge in the entire input sequence, not just those of $\sigma^j$.

Suppose that when an edge $e_i = (u_i, v_i)$ arrives, $u_i$ and $v_i$ have the same color. This can only happen if they have the same level $j$. We first check if appending $e_i$ to $\sigma^j$ is still moderate. If it is, then we append and send $e_i$ to $\alg_j$, and recolor vertices of level $j$ accordingly. Otherwise, we choose one of them, arbitrarily, to promote to level $j+1$. 

When we promote a vertex $u$ to level $k$, in order to maintain invariant (2), we attempt to append, one-by-one, each edge $(u,v)$ in which $\Part(v)=k$ to $\sigma^k$, in arbitrary order. In this process, if for some edge $(u,v)$, we are unable to append it to $\sigma^k$ without maintaining the moderate property, we promote $u$ to the next level. Note a vertex cannot be promoted indefinitely as it eventually ends up at a level with no other vertices.
Pseudocode of our algorithm is given below.

\noindent\fbox{\vbox{\small
\vspace{1mm}
\noindent
\textbf{Algorithm $\cC$: Coloring a bipartite graph of largest bond size $\beta$ using
  $O(\epsilon \log n)$ colors.}%
\begin{enumerate}[\hspace{1em}a.]
\item Set $\Part(v):=1$ for all $v\in V$.
\item Set $c^1$ to be the initial vertex coloring.
\item Fix an arbitrary initial coloring $c^j$ for $j > 1$, say $c^j(v)=2j$ for all
$v\in V$.
\item\label{st-tau}
Let $\tau=\max\Set{2^{1/\epsilon},4\beta^2/\alpha}$. 
\hfill\textit{// $\alpha$ is a constant parameter in $(0,1)$}
\item When edge $e_i = (u_i,v_i)$ arrives:
\begin{enumerate}[\hspace{1em}(1)]
  \item If $\Part(v_i)\ne\Part(u_i)$ then return;\hfill\textit{// different
      colors}
  \item  Let $j=\Part(v_i)$ \hfill\textit{// $j=\Part(u_i)$ as well}
  \item If $\sigma^j \circ (e_i)$ is
    $(\tau,\alpha)$-moderate, then  send $e_i$ to $\alg_j$ \hfill\textit{// $e_i$ is
      appended to $\sigma^j$}
  \item\label{st-pr} Else invoke $\promote(u_i, j+1)$
    \hfill\textit{// $e_i$ is $j$-excess; $u_i$ is chosen arbitrarily}
\end{enumerate}
\item Output coloring $c(v)=c^{\Part(v)}(v)$ for all $v\in V$.
\end{enumerate}

\noindent\textbf{Procedure $\promote(u,k)$:}
\begin{enumerate}
\item Let $V':=V^k\cup\Set{u}$
  \hfill\textit{// $u$ gets the initial color of $c^k$}
  \item Let $\sigma'$ be an arbitrary ordering of the edges $e=(u,v)$ arrived so far, such that $\Part(v) = k$. 
\item For each edge $e'$ in $\sigma'$:
\begin{enumerate}
    \item \label{step-pr-rec}
    If $\sigma^k\circ(e')$  is
    $(\tau,\alpha)$-moderate, send $e'$ to $\alg_k$ 
     \hfill\textit{// $e'$ is appended to $\sigma^k$}  
    \item \label{pr-pr} Else invoke $\promote(u, k+1)$ and return
    \hfill\textit{// $e'$ is $k$-excess}
\end{enumerate}
\item $\Part(u):=k$
\hfill\textit{// promotion to level $j$ was successful}
\end{enumerate}
}
}

\subsection{Analysis}

\begin{claim}
Let $e_i=(u,v)$. At any time $t\geq i$,  if $\Part(u)=\Part(v)=k$, then $e_i \in \sigma^k$.
\end{claim}
\Proof
At any time $t \geq i$, if at the beginning of time step $t$, $\Part(u)\neq\Part(v)$, we are done as the levels do not change (step 5(a)).
 If at the beginning of  time step $t$ 
$\Part(u) = \Part(v)= k $ and $\sigma^k \circ e_i$ is moderate then $e_i$ is added $\sigma^k$ (step 5(c)), and the claim holds.
In the remaining case (Step 5(d)), $u$ is promoted and $v$ is not, hence their level will not be the same at the end of time step $t$.
\QED

The correctness of the algorithm is now straightforward. 
\begin{lemma}
  \label{lem:algb-correct}
  The coloring produced by Algorithm $\cC$ is correct. 
\end{lemma}
\Proof
Follows from the fact that at any time  (1) two vertices of different levels are not colored by the same color, (2) if an edge $e=(u,v)$ exists and $\Part(u) = \Part(v)= k $, then $e\in \sigma^k$ (3) the correctness of  $\alg_k$ for all $k$.
\QED

We now proceed to give an upper bound on the cost paid by Algorithm $\cC$
and on the number of colors it uses. 
Recoloring occurs in $\cC$ either by some $\alg_j$ or by procedure
$\promote$ (which recolors to the initial color of the corresponding level).
We analyze each separately.

Let us denote by $\Opt_2(\sigma)$ the optimal cost of recoloring an input sequence $\sigma$, when only the two initial colors of unit cost are available; the initial coloring of the vertices is understood from the context. Regarding recoloring by some $\alg_j$, we have the following. Let $R_j$  denote the set of
vertices whose color was altered by $\alg_j$.

\begin{lemma}
  \label{lem:nice-ratio}
  For all $j$, 
  $\Cost_\alg^j(\sigma^j)= O\left(|R_j|\log \tau /(1-\alpha) \right)$.
\end{lemma}

\Proof
Follows from \lemmaref{lem:moderate} and the fact that $\sigma^j$ is $(\tau,\alpha)$-moderate by construction.
\QED

We note that the recoloring done by the algorithm are either (1) changes of $c^k(v)$, for $v$ such that $\Part(v)=k$, done by $A_k$, or (2) change of the color of $v$ due to the change of $\Part(v)$, in procedure $\promote$. We now analyze the cost due to the latter; the former was analyzed in the above lemma.

 We use the following
definition.
\begin{definition}
    An input edge $e=(u,v)$ is called \emph{$j$-excess} if it 
    caused $\promote(u,j+1)$ to be invoked, 
    i.e., either of the two lines (1) \stepref{st-pr} or (2) \stepref{pr-pr} of Procedure $\promote$, was reached when $e$ was processed.   
    We denote the set of all $j$-excess edges by $F_j$.
\end{definition}

Clearly, the total
cost due to recoloring by $\promote$ is at most $\sum_j |F_j|$, because
each excess edge causes at most one vertex to be recolored (indirectly)
when $\promote$ changes the level of a vertex. 
  Then by Lemma~\ref{le:special_edges} we have:

\begin{lemma}
 \label{lem:Fj}
   $|F_j| \leq \beta\frac{|R_j|}{\alpha\tau}$ for every $j\ge1$.
 \end{lemma}

We now have that the total cost of the algorithm is 
\begin{equation}
\label{eq:alg-cost}
    \Cost_\cC(\sigma)
  \leq \sum_{j\ge2} |F_j| + \sum_{j\ge1}\Cost_\alg(\sigma^j)
  \leq O\left(\frac{\beta}{\alpha\tau} + \frac{\log \tau}{1-\alpha}\right)\sum_{j \ge 1}|R_j|.
\end{equation}
Next, we bound $\sum_{j \ge 1}|R_j|$. We will need the following lemma.

\begin{lemma}
  \label{lem:Ej}
  Let $E^j$ denote the set of edges in $\sigma^j$. Then for every $j>1$, 
  $|E^j| \leq \beta^2\cdot\frac{|R_{j-1}|}{\alpha\tau}$.
\end{lemma}

\Proof Let $S_j$ be the set of vertices $u$ to which  $\promote(u,j)$ was
applied at some point in the algorithm. Clearly, $|S_j|=|F_{j-1}|$, because 
each $(j-1)$-excess edge promotes a single vertex to level $j$.
It follows that $|S_j|=|F_{j-1}| \leq 
\beta\cdot\frac{|R_{j-1}|}{\alpha\tau}$ by
\lemmaref{lem:Fj}.
Now, 
$E^j \subseteq G[S_j]$, because an edge is appended to $\sigma^j$ only if
both its endpoints are in $V^j$ at that time. Since $E^j$ is a subset 
of the edges of a graph with largest bond size $\beta$, we have from 
\corollaryref{cor:bond} that $|E^j| \leq \beta(|S_j|-1)$.
We therefore conclude that $|E^j| \leq \beta^2\cdot\frac{|R_{j-1}|}{\alpha\tau}$.
\QED

We can now prove  that  $|R_j|$ decreases exponentially
with $j$, for large enough $\tau$. 
\begin{proposition}
\label{prop:Rj}
    $|R_j| \leq 2\beta^2\cdot\frac{|R_{j-1}|}{\alpha\tau}$ for every $j > 1$.
\end{proposition}

\Proof Since $\alg_j$ recolors a vertex only if it is incident on an
input edge, we have that
$|R_j| \leq 2|E^j|$. The result follows from \lemmaref{lem:Ej}.
\QED

\propositionref{prop:Rj} immediately gives an upper bound on the number
of colors used by Algorithm $\cC$. %
\begin{corollary}
\label{cor:colors}
With $\alpha = \frac{1}{2}$ and $1/\epsilon\ge2\log\beta+3$,
Algorithm $\cC$ uses  at most $O(\epsilon^{-1}\log n)$ colors.
\end{corollary}
\Proof 
Let $\gamma=\frac{2\beta^2}{\alpha\tau}$. By \stepref{st-tau} of 
Algorithm $\cC$, $\gamma<1$.
Clearly, $|R_1|\le n$, and hence, by \propositionref{prop:Rj} we have
that $|R_j|\le n\cdot\gamma^j$. 
Hence, for $j>\log_{(1/\gamma)}n$ we
have that $|R_j|<1$. The result follows, since
$\log(1/\gamma)=\log\tau-O(\log\beta)=\Theta(1/\epsilon)$. 
\QED

We can now conclude the analysis of Algorithm $\cC$.

\ProofOf{Theorem}{thm:tradeoff}
Fix $\alpha=1/2$.
The number of colors is given by \corollaryref{cor:colors}. By Inequality~\eqref{eq:alg-cost}, the total cost of the algorithm is at most 
\[O\left(\frac{\beta}{\alpha\tau} + \frac{\log \tau}{1-\alpha}\right)\sum_{j \ge 1}|R_j|.\]

Since $\epsilon > \frac{1}{2 \log \beta + 3}$ and $\alpha = 1/2$, we get that $\tau = 2^{1/\epsilon}$ and so $\frac{\beta}{\alpha\tau} + \frac{\log \tau}{1-\alpha} \leq O(1/\epsilon)$. Moreover, Proposition~\ref{prop:Rj} implies that $\sum_{j \ge 1}|R_j| \leq O(1)|R_1|$. By Lemma~\ref{lem:size}, we have that $|R_1| \leq O(1)\Opt_2(\sigma^1) \leq O(1)\Opt_2(\sigma)$. Thus, $\sum_{j\ge1}|R_j| \leq O(1)\Opt_2(\sigma)$. Therefore,  the total cost of the algorithm is at most $O(1/\epsilon)\Opt_2(\sigma)$, as desired.
\QED

\section{Lower Bounds}
\label{sec:lb}
In this section we prove two lower bounds on online recoloring.
The first, \theoremref{thm:lb-paths}, says that no matter how many extra
colors we use, if they cost $D$, then the best competitive ratio one
can hope for is $O(\log D)$ even if the underlying graph is acyclic.
The second lower bound, \theoremref{thm:lb-bond}, shows that
if there is no upper bound on the bond size of the graph, the competitive
ratio of any online recoloring algorithm is at least $\Omega(\min(\log n, D))$.

\subsection{Acyclic Graphs}
\label{sec-lb}
In this subsection we show that the competitive ratio of Algorithm
$\cB$ cannot be improved, even by randomized algorithms, as stated
by \theoremref{thm:lb-paths} reproduced below. 

\thmLBpaths*

\Proof
By Yao's Principle, it suffices to show the existence of an input
distribution on which 
the expected cost of any deterministic algorithm is $\Omega(\log D)$
times the optimal cost.  We describe such a distribution. Initially,
there are $n$ vertices, each colored by one of the two basic (unit cost)
colors.  Without loss of generality, let us assume that  $n$ is a
multiple of $\ge D/\log D$ vertices. For ease of explanation, we shall assume that the vertices are arranged left-to-right on a line. The input sequence is 
constructed in $H$ phases, where
$H\DEF\lceil \log D - \log \log D \rceil$. In phase $1\le h\le H$, edges
arrive so that at the end of the phase, each connected component is a
path with $2^h$ vertices. Specifically, for each $1 \leq j \leq n/2^h$, an edge connecting random endpoints of the $(2j-1)$-th leftmost path and the $2j$-th leftmost path is inserted.
Note that after the last phase, we have a set of paths of length $2^H$ vertices
each. Note that for every phase $h$, the set of connected components at the end of the phase is predetermined; it is only the order in which the vertices in a connected component appear in the corresponding path that is randomized.

Fix an arbitrary deterministic recoloring algorithm $A$. We now bound
from below the expected cost of $A$ on the input distribution defined
above. We first prove the following lemma.

\begin{lemma}
  \label{lem:fraction}
  Let $A^*$ be an arbitrary deterministic algorithm that uses only the
  two basic colors.  Let $k$ be any integer power of $2$ not greater
  than $2^H$.  Consider a randomized input sequence as described above and a connected component consisting of $k$ vertices after $\log k$ phases. Let $\sigma$ be the restriction of the input sequence to the connected component. Then we have
  $\Pr\left[\Cost_{A^*}(\sigma) \geq k (\log k - 2)/8\right] \geq 1/2$.
\end{lemma}

\Proof
In phase $h >1$, there are $k/2^{h}$ merges of paths each
containing $2^{h-1}$ vertices. Observe that with probability exactly
$1/2$ a given merge is monochromatic (i.e., the merging new edge
connects two vertices of the same color)
independently of all other merges (in the same and other phases).
This is because for a given merge, each of the two merged paths have
one end colored with each of the two basic colors. Further observe
that a monochromatic merge at phase $h>0$ results in recoloring of
exactly $2^{h-1}$ vertices.  Let $Z_h$ be the indicator random variable
indicating whether at least half of the merges in phase $h$ are
monochromatic.
Clearly, if
$Z_h = 1$ then the algorithm recolors at least
$\frac{1}{2}\frac{k}{2^h}\cdot 2^{h-1} = k/4$ vertices in phase $h$ .
Moreover, for any $1<h<\log k$, even
when conditioned on all $Z_j$ (except $Z_h$), it holds that
$\Pr[Z_h = 1] = 1/2$.

It therefore follows that
$\Cost_{A'} \geq \sum_{h = 2}^{\log k} Z_h \cdot k/4$, and 
     \[\Pr\left[\Cost_{A^*} \geq \frac{k (\log k - 2)}{8}\right] \geq \Pr\left[\sum_{h = 2}^{\log k} Z_h \cdot \frac{k}{4} \geq \frac{k (\log k - 2)}{8}\right] \geq \frac{1}{2},\]
    where the last inequality follows from the fact that 
    $\sum_{h=2}^{\log k-1}Z_h$ is a sum of $(\log k - 2)$ 
    independent Bernoulli random variables with parameter $1/2$,
    and so their sum is 
    at least $(\log k - 2)/2$ with probability $1/2$.
\QED

We now continue with the proof of \theoremref{thm:lb-paths}.
 Consider one of the paths of length $2^H$ that exist after the
last phase. We give a lower bound on the expected cost of $A$ for
re-coloring the vertices of that path.  Let $A_D$ be the indicator random
variable indicating if $A$ uses at least once an extra color (of cost
$D$) for recoloring the vertices of the given path.  There are two
cases to consider.
\begin{enumerate}[(1)]
\item If $\Pr\left[A_D\right] \geq 1/4$,
  then $\Pr\left[\Cost_{A} \geq D\right] \geq 1/4$,
  and hence $E[\Cost_{A}] \geq D/4$. 
\item Otherwise, $\Pr\left[A_D\right] < 1/4$. In this case, define a
  deterministic online recoloring algorithm $A'$ that uses only the
  basic colors as follows.
  $A'$ mimics  $A$ so
  long as $A$ uses basic colors. Once $A$ uses a special
  color (if it does), $A'$ stops mimicking $A$ and
  continues to recolor (properly), using only the basic colors.
\end{enumerate}
We make the following observations.
\begin{itemize}
\item $\Pr\left[\Cost_{A'} \geq  2^H (H - 2)/8\right]  \geq 1/2$ by
  \lemmaref{lem:fraction}.
\item The probability that $A'$ and $A$ diverge in their colorings on the path under consideration   (and hence {\em do not have} the same cost for that path) is less than $1/4$.
\end{itemize}
It follows that with probability more than $1/4$,
$\Cost_A=\Cost_{A'}$, and that cost is at least
\begin{align*}
\Cost_A~\ge~& 2^H \cdot\frac{H - 2}{8}\\
~=~&  (2^{\ceil{\log D - \log \log D}}\cdot \frac{\ceil{\log D - \log\log D}-
  2}8\\
~\ge~&\frac{D}{\log D}\cdot\frac{\log D-\log\log D-2}{8}\\
~=~ &\Omega(D)~.
\end{align*}
Therefore, in both cases the expected cost of
$A$ for recolorings of the vertices in the path that we consider is
$\Omega(D)$.

The bound above is the cost for a single component of length $2^H$.
By linearity of expectation, the expected cost of $A$ for all components
is $\frac{n}{2^H}\cdot \Omega(D)$. On the other hand the cost of
the optimal algorithms is $O(n)$ for any input sequence.  Hence we
have a lower bound on the competitive ratio of
$\Omega\left(\frac{(n/2^H) \cdot D}{n}\right)
=\Omega\left(\frac{D}{2^H}\right)
=\Omega\left(\frac{D}{2^{\log D - \log \log D}}\right)
=\Omega(\log D)$.
\QED

\subsection{Bipartite Graphs with Unbounded Bond Size}
In this subsection we show that no deterministic online recoloring
algorithm can have good competitive ratio for all  bipartite graphs with
large bond,  even when special colors are available.
In fact, using
special colors cannot improve the 
competitive ratio  beyond  the easy $O(D)$ bound of
\lemmaref{lem:cr=D} if the bond may be large. %

\thmBond*

\Proof 
We construct an adversarial input sequence that consists of phases.
At any point in time, a vertex is said to be {\em special} if the
algorithm has recolored it at some earlier point to a special color; the
vertex is called {\em basic} otherwise. Furthermore, at each phase we
characterize the connected components as {\em active} or {\em
  inactive}. A component is said to be active if at least half of its
vertices are basic, and it is said to be inactive otherwise.  Let
$c_1,c_2$ denote the two basic colors. Initially, each vertex is given
either the color $c_1$ or $c_2$.  An active component is said to be
{\em $c$-dominated}, for $c\in\Set{c_1,c_2}$ if at least a quarter of its
vertices are colored $c$ (a component may be both $c_1$-dominated and
$c_2$-dominated). 

The construction maintains the following invariant:
\begin{itemize}
\item At the start of phase $h>0$, every active component has
  $2^{h-1}$ vertices.
\end{itemize}
In phase $h$, we match
active components of the same dominating color in pairs.
For each such pair $(C, C')$ of connected components of the same
dominating color,
we add a perfect matching between their vertex sets such that at least
a quarter  of the matching edges (i.e., at least $2^{h-3}$ edges) are
monochromatic. The process ends once there is at most one active
component of dominating color $c_1$ and at most one active component
of dominating color $c_2$.

We claim that the cost  paid by the algorithm for
the above input sequence  is at least
$\Omega(\min\{n\log n, nD\})$.
To prove the claim, let $k$ be the number of special vertices at the end of the input
sequence, and let $M$ be the number of edges that were monochromatic
when they arrived. Clearly, the cost for the algorithm is at least
$\max\Set{kD,M}\ge (kD + M)/2$. %
Now, if $k \geq n/4$, %
then the algorithm's cost is $\Omega(nD)$ and the claim holds.
Otherwise, $k < n/4$, and hence, by the definition of active
components, at the end of the execution, there are at most $2k < n/2$
vertices in inactive components. It follows that in each phase there
are at least $n/2$ vertices in active components, and therefore at
least $n/4$ vertices in active components of the same dominating
color, which implies that in each phase, the number of introduced
monochromatic edges is $\Omega(n)$. Further, note that if $k<n/4$ then
the number of phases is at least $\log n-3$: before every phase
$h \le \log n - 2$, each connected component contains less than
$2^{\log n -3} = n/8$ vertices, and hence there are at least $4$
active connected components, so there are at least two connected
components dominated by the same color. It follows that the total cost
of the algorithm is $\Omega(n \log n)$.

In summary, we have shown that the cost of the online algorithm for the input sequence
specified above is
$\Omega(\min\{n\log n, nD\})$. As the optimal cost is $O(n)$
the theorem follows.  \QED

We note that the largest bond size of any active connected component $C$ in the
construction above is $\Omega(|C|)$.

 \begin{toappendix} 
 \section{Algorithm for Non-Uniform Cost with \(\Delta\) special colors}
\label{sec:A-non-uniform}

In this section, we give a variant of the algorithm for non-uniform costs, which works with only $\Delta$ special colors, at the cost of a somewhat more complicated algorithm and proof. We use some of the definitions and notations of Section~\ref{sec:non-uniform}. 
The basic strategy of the algorithm is the same as that of the algorithm given in Section~\ref{sec:non-uniform}; however the treatment of the special vertices is somewhat different (since we have only $\Delta$ special colors, and the degree of a special node could be $\Delta$). This is reflected in a different procedure $\recx$, the rest of the algorithm is the same.

The algorithm, called $\hat\cB$, is presented in \figref{fig:hatB}.
Its analysis is presented in \sectionref{sec:hatB-cor} and \sectionref{sec:hatB-comp}.

\begin{figure}
 \noindent
 \hspace{-1mm}\fbox{\vbox{\small
 \vspace{1mm}
 \noindent
 \textbf{Algorithm $\hat{\cB}$}
 \vspace{2mm}\\
 \textbf{State:}
 \begin{itemize}\vspace{-1mm}
 \item Each vertex $u$ has an \emph{actual} color $c(u)$
   and a \emph{simulated} color
   $\C(u)$. The initial actual colors are given as input, and the
   initial simulated colors are the initial actual colors.
 \item Each vertex records whether its color was ever changed by 
 the simulation of $\alg$.
 This
 allows the algorithm to maintain the set $R_i$.
 \item Each vertex has an indication whether it is ``special" or not. Initially all vertices are not special.
 \end{itemize}

\noindent
\textbf{Action:}\\
Upon the arrival of edge $e_i = (u_i, v_i)$:%
\begin{enumerate}[\hspace{1em}a.]
\item 
If $\sigma^{\rm{sim}}_{i-1}\circ e_i$  is
    $(D,\alpha)$-moderate then \\
    \hspace*{5mm} send $e_i$ to \Alg   (which updates $\C(\cdot)$);
    \hfill \textit{//$\sigma^{\rm{sim}}_{i}= \sigma^{\rm{sim}}_{i-1}\circ e_i$ } \\
    \hspace*{5mm} set $c(w):=\C(w)$ for  every non-special vertex $w$.
\item Else  \label{line:A-mark}\\
\hspace*{5mm}If both $u_i$ and $v_i$ are not special then mark $u_i$ as {\em special}.
\item   
\label {st:A-special}
 Invoke $\recx(u_i,v_i)$
\label{st:A-simulate}
\end{enumerate}

 \noindent
 \textbf{Procedure} $\recx(u,v)$:%
 \begin{enumerate}%
\item If $u$  and $v$ are special  then \\
\hspace*{5mm} if $e=(u,v)$ is monochromatic then $\recolor(u)$
\item  Else \\
\hspace*{5mm}   If $u$ is special then \\
\hspace*{10mm}  if $u$  is colored by a basic color then $\recolor(u)$
\item \hspace*{5mm}  Else \\
 \hspace*{10mm}  If $v$ is special then \\
 \hspace*{15mm}  if $v$  is colored by a basic color then  $\recolor(v)$
 \end{enumerate}

 \noindent
 \textbf{Procedure} $\recolor(w)$:%
 \begin{enumerate}[\hspace{1em}i.]
 \item Let $d^*(w)$ be the number of special
   neighbors $w$ has (including the newly-arrived edge). 
 \item \label{A-r2} If $d^*(w)<\Delta$, recolor $w$ using a free   special color.
   \label{recx:A-small}
 \item \label{A-r1} If $d^*(w)=\Delta$, recolor $w$ using a free
   special color if it exists, otherwise use a free basic
   color.
   \label{recx:A-delta}
 \end{enumerate}
}}
\caption{The algorithm uses $\Delta$ additional colors}
\label{fig:hatB}
\end{figure}

\subsection{Correctness}
\label{sec:hatB-cor}
We start by  showing that Algorithm $\hat{\cB}$ produces a valid coloring.  
The following fact is clear from the code.
\begin{fact}
\label{fact:A-special_color}
A vertex colored by a special color is marked special.
\end{fact}
\begin{lemma}
  \label{lem:A-special}
  At any time, if $u$ is a special vertex  colored by a basic color, then  its degree is
  $\Delta$ and all its neighbors are colored by special colors.
\end{lemma}
\Proof
The claim is proved by induction on time steps, the base case is the time step when $u$ is marked special.  This can happen in Line~\ref{line:A-mark} only, when an edge $e_i=(u,v)$ arrives and both $u$ and $v$ are not special. In that case the newly-marked vertex is re-colored by procedures $\recx$ and $\recolor$. By the code of $\recolor$ it  is recolored to a special color unless its (new) degree is    $\Delta$, and there is no free special color (Line~\ref{recx:A-small} and Line~\ref{recx:A-delta}).

For time $j>i$, by the induction hypothesis the claim holds for time $j-1$. 
If $u$ is colored by a basic color at time $j-1$, then we have by the induction hypothesis 
that its degree at time $j-1$ is $\Delta$ 
and all its neighbors are colored by special colors. The claim could be violated at step $j$ only if a neighbor of $u$, say $q$, would be colored by a basic color. But since $q$ always has a free special color (since $u$ is colored by a basic color) this cannot happen by the code of $\recolor$. 

If $u$ is colored by a special color at time $j-1$, then by the code of $\recolor$ it can be colored by a basic color at step $j$ only if 
its (new) degree is    $\Delta$, and there is no free special color (Line~\ref{recx:A-small} and Line~\ref{recx:A-delta}).
\QED

The next lemma is analogous to the respective lemma for Algorithm ${\cal B}$, and its proof is very similar. For completeness we give the full proof below.
\begin{lemma}
  \label{lem:A-valid}
  After every step $i$, $c_i$ is a valid coloring of the graph $G_i=(V,E_i)$.
\end{lemma}
\Proof
To prove the lemma we proceed by induction on $i$. The base case is $i=0$, in which $E_0=\emptyset$ and
hence any coloring is valid. For the inductive step, we first consider all edges but the new edge $e_i$ and for those we proceed in two  sub-steps. Then we consider the new edge $e_i$.

For all edges but $e_i$,
if $e_i$ is not a simulated edge and is not sent to $\cA$, then the first 
sub-step is empty.
If $e_i$ is  a simulated edge and is sent to $\cA$, then the colors of some non-special vertices may change.
After this sub-step, for each (old) edge: (1)  if  its two endpoints are non-special then the edge is not monochromatic by the correctness of $\cA$; (2) if its two endpoints are special,  then  their color does not change in this sub-step and the edge is not monochromatic by the induction hypothesis; (3) if one endpoint is special and the other is not, then by Lemma~\ref{lem:A-special} the special endpoint is colored by a special color
(and the non-special endpoint obviously by a basic color) and hence the edge is not monochromatic.

The second sub-step is the invocation of $\recx$ on the input edge. Following this invocation one  (special) vertex might be recolored. Any edge that is not adjacent to the vertex that was recolored remains non-monochromatic by the induction hypothesis. For an edge adjacent to the node that is recolored, then the code of
$\recolor$ ensures that the edge is not monochromatic.

Now as to the edge $e_i$ itself, we have a number of cases depending whether it is a simulated edge and the status (special or not) of its two endpoints when step $i$ begins.
\begin{enumerate}
\item If its two endpoints are non-special when step $i$ starts:
\begin{itemize}
\item if $e_i$ is a simulated edge: at the end of the first sub-step $e_i$ is not monochromatic by the correctness of $\cA$; none of its endpoints is marked special and hence $\recx$ does not recolor any node and $e_i$ remains non-monochromatic.
\item  if $e_i$ is not a simulated edge: One of its endpoints is marked special; $\recx$ recolors that vertex and by the code of $\recolor$ $e_i$ is not monochromatic.
\end{itemize}
\item  If its two endpoints are special when step $i$ starts, then  regardless of whether $e_i$ is a simulated edge or not the code of $\recx$ ensures that if $e_i$ is  monochromatic when step $i$ starts, then one of the endpoints of $e_i$ is recolored by $\recolor$ whose code ensures that $e_i$ is not monochromatic when step $i$ ends.
\item If one endpoint (w.l.o.g. $u$)  is special and the other ($v$) is not  when step $i$ starts, then regardless  of whether $e_i$ is a simulated edge or not the status of neither vertices change. If $u$ is colored by a special color  when step $i$ starts (and $v$ obviously by a basic one) then $u$ is not recolored by $\recx$ and edge $e_i$ is non-monochromatic. If  $u$ is colored by a basic color when step $i$ starts, then $u$ is recoloredby $\recx$, but the code of $\recolor$ ensures that $e_i$ is not monochromatic.\QED
\end{enumerate}

\subsection{Competitive Analysis}
\label{sec:hatB-comp}

To bound from above the cost of Algorithm $\hat{\cB}$, we note that Algorithm $\hat{\cB}$ is the same as 
Algorithm $\cB$ except for procedure $\recolor$. That procedure sometime recolors a special vertex by a basic color, incurring a cost of $1$ instead of at most $D$, while for Algorithm $\cB$ all recolorings of special vertices are by special colors incurring a cost of $D$. The cost of Algorithm $\hat{\cB}$ is therefore upper bounded by that of Algorithm $\cB$, for the same input sequence.
We can therefore conclude with the following theorem, which is 
is an analogue  of Theorem~\ref{thm:cost-D}.

\begin{theorem}
  \label{thm:A-cost-D}
Suppose that the final graph is bipartite, with all vertex degrees at most
$\Delta$,  and with bond size at most $\beta$ for each  of its
connected components.  Then Algorithm $\hat{\cB}$ uses $\Delta$ special colors of cost at most $D$, and has
  competitive ratio  $O(\log D+\beta^2)$.
\end{theorem}

 \end{toappendix}

\section{Conclusions}
\label{sec:conc}
In this paper we studied competitive vertex recoloring with weighted
augmentation. We have shown that the competitive ratio
of recoloring bipartite graphs may be reduced if the online
algorithm can use additional colors, even if they are more costly
than the basic two colors. Beyond the direct technical contribution, 
we believe that we introduced some conceptual contributions:
\begin{itemize}
\item The approach of weighted resource augmentation is new, 
as far as we know.
\item The concept of the largest graph bond as a way to generalize  
algorithmic results quantitatively.
\end{itemize}
It seems that these ideas can be useful in dealing with other 
problems of competitive analysis.

\bibliography{./recoloring}

\end{document}